\documentclass[preprint,amsmath,amssymb,aps,a4]{revtex4}
\usepackage{xcolor}
\usepackage{graphicx}
\usepackage{amssymb}
\usepackage{multirow}

\newcommand{\bfz}{{\bf 0}_{\perp}}

\newcommand{\bfk}{{\bf k}_{\perp}}

\newcommand{\Dp}{{\bf \Delta}_{\perp}}

\usepackage{xcolor}
\begin{document}
\title{Chiral-odd generalized parton distributions for the low-lying octet baryons}
	
\author{Navpreet Kaur}
\email{knavpreet.hep@gmail.com}
\affiliation{Department of Physics, Dr. B.R. Ambedkar National
		Institute of Technology, Jalandhar, 144008, India}
	
\author{Harleen Dahiya}
\email{dahiyah@nitj.ac.in}
\affiliation{Department of Physics, Dr. B.R. Ambedkar National
		Institute of Technology, Jalandhar, 144008, India}
	
\date{\today}%
\begin{abstract}
We have studied the chiral-odd generalized parton distributions (GPDs) for the octet baryons within the framework of diquark spectator model, emphasizing the difference arising from different quark flavors of $p$, $\Sigma^+$ and $\Xi^o$. The quark-quark correlator for the tensor current has been solved to investigate the transversely polarized quark dynamics using light-cone wave functions. In the forward limit, transversity distributions of all the constituent quark flavors of $p$, $\Sigma^+$, and $\Xi^o$ have been studied. Additionally, the lowest moment of GPD $H_T$ and the combination of GPDs $E_T$ and $\tilde{H}_T$ have also been investigated. Our results are comparable with other available model predictions and lattice data for the case of $p$. Furthermore, we have presented a comparative analysis of tensor charges and anomalous tensor magnetic moments for all the constituent quark flavors of the considered baryons, comparing them with available data obtained from different approaches.
\end{abstract}
%
\maketitle
\section{Introduction}
Parton distributions provide a fundamental description of the hadron structure in terms of partons which are the elementary degrees of freedom in quantum chromodynamics (QCD). At the level of leading twist, quark distributions portray the basic quark structure of a hadron. The unpolarized distribution $f_1(x)$, regardless of the spin orientation, corresponds to the probability of locating a quark with $x$ longitudinal momentum fraction of the parent hadron. In a longitudinally polarized hadron, the net helicity of a quark is measured by the helicity distribution $g_1(x)$. Both of these distributions are well characterized because of their chiral-even nature and can be accessed through deep inelastic scattering (DIS). The distribution $h_1(x)$ represents the difference between the number densities of quarks whose polarization aligns with the hadron and those with opposite polarization in a transversely polarized hadron and thus, termed as transversity distribution. It is opposite to the tensor current of the hadron and being chiral-odd in nature, its experimental measurement become difficult. Since the tensor structure is not directly measurable, its coupling with another chiral-odd entity becomes essential within the scattering process to measure it. The azimuthal single spin asymmetry in semi-inclusive deep inelastic scattering (SIDIS) \cite{Artru:1989zv, Anselmino:2007fs}, the transverse spin asymmetry Drell-Yan process in proton-antiproton reactions \cite{Anselmino:2004ki, Efremov:2004qs, Pasquini:2006iv}, selected photoproduction reactions of lepton pairs \cite{Pire:2009ap} and diffractive exclusive production of two vector mesons \cite{Ivanov:2002jj, Enberg:2006he} are considered as valuable processes to gain fruitful insights on the hadron transversity.

The helicity dependent three-dimensional (3D) structure interpretation of a hadron can be interpreted through chiral-odd generalized parton distributions (GPDs) in the momentum space. Depending on the different helicity orientations of the initial and final state of a hadron and constituent quarks, there are four different chiral-odd GPDs $H_T$, $E_T$, $\tilde{H}_T$ and $\tilde{E}_T$. The last GPD vanishes due to the time-reversal invariance \cite{Diehl:2001pm} and we are left with only three chiral-odd GPDs for the case of zero skewness. These GPDs depend on the momentum transferred in the process along with the average longitudinal momentum fraction $x$. In the forward limit, when momentum transfer becomes zero, the GPD $H_T$ reduces to transversity distribution $h_1(x)$. The lowest moment of these GPDs correspond to the tensor form factors. Specifically, at zero momentum transfer, the lowest moment of GPD $H_T$ results in the tensor charge, and the combination of GPDs $E_T$ and $\tilde{H}_T$ leads to the anomalous tensor magnetic moment. Experimental results on $h_1(x)$ \cite{Anselmino:2013vqa} have been extracted through the combined SIDIS data from HERMES \cite{HERMES:2010mmo} and COMPASS \cite{COMPASS:2012ozz, Martin:2013eja} by utilizing the Collins fragmentation function reported by Belle collaboration \cite{CDF:2012cxh}. They have also measured the values of tensor charge for the constituent quark flavors of proton and reported it as $0.39_{-0.12}^{+18}$ and $-0.25_{-0.10}^{+30}$ for $u$ and $d$ quark flavors respectively \cite{Anselmino:2013vqa}. Based on the investigation of pion-pair production in DIS of transversely polarized targets, the valence quark transversity distributions and tensor charges have been extracted \cite{Bacchetta:2012ty}  using the released data from the HERMES \cite{HERMES:2008mcr} and COMPASS \cite{COMPASS:2012bfl} collaborations, combined with the Belle data of $e^+e^-$ annihilations \cite{Belle:2011cur}. In Ref. \cite{Kang:2015msa}, the valence quark contributions to the nucleon tensor charge were also estimated, along with their transversity distribution, by studying the transverse momentum-dependent evolution of the Collins azimuthal asymmetries in $e^+e^-$ annihilations and semi-inclusive hadron production in DIS, including the approximate next-to-leading-logarithmic order. Lattice calculations have also investigated the valence quark transversity distributions and tensor charges in Refs. \cite{Lin:2017stx, Chen:2016utp} for proton. In a relativistic quark model \cite{Schmidt:1997vm} and Schwinger-Dyson formalism \cite{Yamanaka:2013zoa, Pitschmann:2014jxa}, nucleon tensor charges have also been predicted.

Transversity distribution accounts for only one-dimensional information. To interpret the 3D structure, chiral-odd GPDs for the case of proton have been studied in constituent quark model \cite{Pasquini:2005dk}, lattice QCD \cite{Gockeler:2005cj}, quark-diquark model (inspired from the soft-wall AdS/QCD) \cite{Maji:2017ill}, basis light-front quantization \cite{Kaur:2023lun} and bag model \cite{Tezgin:2024tfh}. 
Extensive work has been conducted on the case of proton. In contrast, comparatively less work has been reported for hyperons in the chiral-odd sector despite their similar spin and parity. The tensor form factors of the octet baryons are computed in the self-consistent SU(3) chiral quark-soliton model \cite{Ledwig:2010tu}. Leading order light-cone QCD sum rules have been used to study the transition tensor form factors, along with the distribution amplitudes of $\Sigma$, $\Xi$, and $\Lambda$ hyperons \cite{kucukarslan:2016xhx}. Besides this, the electric and magnetic form factors of the octet baryons have been explored by using lattice QCD (chiral extrapolation) \cite{Shanahan:2014cga, CSSM:2014knt} and  a covariant quark model \cite{Ramalho:2024wxp, Ramalho:2019koj}. Chiral-even GPDs with and without skewness parameter, electromagnetic form factors, and charge densities have been explored for the case of low-lying octet baryons \cite{Kaur:2023zhn, Kaur:2024yqr}. The leading-twist light-cone distribution amplitudes as well as quasi-distribution amplitudes for octet and decuplet baryons within large momentum effective theory have also been studied \cite{Han:2024ucv}. In Refs. \cite{Dahiya:2002qj, Dahiya:2002fp}, octet and decuplet baryon magnetic moments in the chiral quark model have been investigated. In recent years, the elastic form factors of hyperons in the timelike region and their transition form factors have been analyzed at BaBar \cite{BaBar:2007fsu}, CLEO \cite{Dobbs:2014ifa}, Belle \cite{Belle:2022dvb} and BESIII \cite{BESIII:2023ldb, BESIII:2023ioy, BESIII:2021rkn}. Upcoming measurements at PANDA are also set to expand the current dataset \cite{PANDA:2016fbp}. The study of hyperon dynamics provides valuable contributions to hypernuclear physics \cite{Feliciello:2015dua} and plays a crucial role in exploring neutron and strange stars where hyperons populate the inner core at high baryonic densities \cite{Wang:2005vg}. By incorporating effective hyperon and its constituent quark masses, along with the effective momentum into the vacuum quark distributions, the influence of the dense medium on a hyperon in these stars can be investigated as predicted for the lightest strange meson, i.e., kaon \cite{Singh:2024lra} and the octet baryons \cite{Ramalho:2025kii, Ramalho:2024tdi}. Drawing motivation from these studies, we have presented the comparative analysis of chiral-odd GPDs, form factors, tensor charges and anomalous tensor magnetic moment among the constituent quark flavors of octet baryons, including proton $p$ and hyperons  such as $\Sigma^+$ and $\Xi^o$. 

In the present calculations, we have used the diquark spectator model based on the light-cone dynamics. The choice of the light-cone framework is driven by its inherent ability to account for relativistic motion and boosts, making it well-suited for studying quark dynamics within hadrons \cite{Dirac:1949cp, Brodsky:2000ii}. In the diquark spectator model, three valence quark flavors of a baryon are treated as an active quark (participating in the process), while the other two form a spectator diquark. In addition to the scalar diquark, we have also assumed the axial-vector diquarks. Further, we have considered the axial-vector diquarks as isoscalar and isovector spectator diquarks to achieve realistic flavor analysis. To derive the light-cone wave functions (LCWFs), we have choosen the dipolar form for the baryon-quark-diquark vertex over the point-like form to prevent divergence. Different choices are available while summing over the complete set of vector diquark polarization states. This study adopts the sum over transverse and longitudinal polarization, skipping an unphysical timelike polarization state. This diquark spectator model has been used successfully to investigate the transverse momentum-dependent parton distributions \cite{Bacchetta:2008af}, fragmentation functions \cite{Jakob:1997wg}, Sivers functions \cite{Bacchetta:2003rz} and flavor dependence of the Boer-Mulders function \cite{Gamberg:2007wm} with different summations over polarization vectors of diquarks.

The paper is organized as follows. In Sec. \ref{SecModel}, the formalism adopted to calculate the chiral-odd GPDs has been discussed for the dipolar baryon-quark-diquark vertex form. In Sec. \ref{secnumpar}, numerical parameters used in the present  calculations have been presented. Chiral-odd GPD have been calculated and discussed in Sec. \ref{SecGPDs}. We have presented the form factors and their comparison with available data in Sec. \ref{secFF}, along with the tensor charges and anomalous tensor magnetic moments. The work has been summarized in Sec. \ref{SecCon}.

\section{Formalism\label{SecModel}} 
\subsection{Kinematics}
In the light-cone framework, we describe a generic four-vector notation for coordinates as $a=[a^+,a^-,a_{\perp}]$. We choose a convenient asymmetric frame in which the respective initial and final state four-vector momentum coordinates of a baryon $X$ are expressed as
\begin{eqnarray}
	P_X&=&\bigg(P_X^+,\frac{M^2_X}{P_X^+},\bf{0}_{\perp}\bigg), \nonumber \\
	P_X^\prime &=& \bigg((1-\zeta)P_X^+,\frac{M^2_X+\Dp^2}{(1-\zeta)P_X^+},-\Dp \bigg) \, ,
\end{eqnarray}
where $M_X$ refers to the baryon mass and $\zeta$ as the skewness parameter \cite{Brodsky:2000xy, Goldstein:2013gra}. The momentum transfer throughout the process in terms of four-vector notation is described by
\begin{eqnarray}
	\Delta=P_X-P_X^\prime=\bigg(\zeta P_X^+,\frac{t+\Dp^2}{\zeta P_X^+}, \Dp\bigg).
\end{eqnarray}
Here, the negative component of  momentum transfer connects $\Dp$, $\zeta$ and $t$ by the following relation
\begin{eqnarray}
	t=2P_X\cdot\Delta=-\frac{\zeta^2M_X^2+\Dp^2}{1-\zeta}.
\end{eqnarray}
In the present calculations, we restrict ourselves to the case of zero skewness ($\zeta=0$) and thus, $t=\Delta^2=-\Dp^2$, which corresponds to the transverse momentum transfer only. The coordinates of the initial and final state of an active quark are respectively expressed by
\begin{eqnarray}
	k=\big(xP_X^+,k^-,\bfk), \nonumber \\
	k^\prime=\big(x^\prime P_X^+,k^{\prime-},\bfk^\prime).
\end{eqnarray}
In the final state, the longitudinal momentum fraction and transverse momentum carried by an
active quark are respectively defined as $x^\prime=x$ and $\bfk^\prime=\bfk-(1-x)\Dp$.

\subsection{Light-cone wave functions for baryons}
For a two body system of a spin-$\frac{1}{2}$ baryon, the instant form SU(6) quark-diquark wave function can be defined as 
\begin{eqnarray} 
|\mathcal{B}\rangle^{\lambda_X} = \cos \theta \sum_q a_q |q_1 ~\mathfrak{s}(q_2 q_3) \rangle ^{\lambda_X} + \sin \theta \sum_{q^\prime} b_q^\prime |q_1^\prime ~\mathfrak{a}(q_2^\prime q_3^\prime) \rangle ^{\lambda_X} \, ,
\label{InstantWfn}
\end{eqnarray} 
where the summation is over different quark-diquark components \cite{Lichtenberg:1968zz}. Fock state $|q_1 ~\mathfrak{s}(q_2 q_3) \rangle$  corresponds to the $q_1$ quark and $\mathfrak{s}(q_2 q_3)$ scalar diquark. Similarly, $|q_1^\prime ~\mathfrak{a}(q_2^\prime q_3^\prime) \rangle$ represents the Fock state of $q_1^\prime$ quark and an axial-vector diquark $\mathfrak{a}(q_2^\prime q_3^\prime)$. The quantity $\lambda_X$ labels the light-cone spin projections of the baryon and can be categorized as $\Uparrow$ or $\Downarrow$. The coefficients $a_q$ and $b_{q^\prime}$ satisfy the normalization condition $\cos^2 \theta \sum_q a_q+\sin^2 \theta \sum_{q^\prime} b_q^\prime=1$. For nucleons and hyperons, belonging to the same octet of spin-$\frac{1}{2}$ baryons, the associated coefficients have different values. The spin-flavor SU(6) symmetry breaking introduces a mixing angle, $\theta$, which is taken as $\pi/4$ in the symmetric case. The instant form wave functions for the $p$ in the form of quark-diquark pair can be expanded as
\begin{eqnarray}
	|p\rangle^{\Uparrow,\Downarrow} &=& \frac{1}{\sqrt{2}} |u ~\mathfrak{s}(ud) \rangle^{\Uparrow,\Downarrow} -\frac{1}{\sqrt{6}} |u ~\mathfrak{a}(ud)\rangle^{\Uparrow,\Downarrow} +\frac{1}{\sqrt{3}} |d  ~\mathfrak{a}(uu)\rangle ^{\Uparrow,\Downarrow},  
\end{eqnarray}
where $\mathfrak{s}$ and $\mathfrak{a}$ denote the scalar and axial-vector diquark respectively. The probabilistic weight distribution among the scalar isoscalar, vector isoscalar and vector isovector configurations is taken as 3:1:2 with respect to the charge-isospin symmetry. For the case of $p$, the flavor decomposition for an arbitrary function can be expressed as
\begin{eqnarray}
f^{p_u} &=& \frac{3}{2} f^\mathfrak{s} + \frac{1}{2} f^\mathfrak{a} \, , \nonumber \\
f^{p_d} &=& f^\mathfrak{a},
\label{FlavorDecomposition}
\end{eqnarray}
which implies that the valence quark distribution of $u$ quark flavor can be interpreted by involving both scalar and axial-vector diquarks, whereas for $d$ quark flavor, the involvement of solely axial-vector diquark is needed \cite{Jakob:1997wg}. In a similar way, the instant form wave functions for the hyperons $\Sigma^+$ and $\Xi^o$ in the form of quark-diquark pair can be expanded as \cite{Lichtenberg:1968zz} 
\begin{eqnarray}
	|\Sigma^+\rangle^{\Uparrow,\Downarrow} &=& \frac{1}{\sqrt{2}} |u ~\mathfrak{s}(us)\rangle^{\Uparrow,\Downarrow} -\frac{1}{\sqrt{6}} |u  ~\mathfrak{a}(us)\rangle^{\Uparrow,\Downarrow} +\frac{1}{\sqrt{3}} |s  ~\mathfrak{a}(uu)\rangle ^{\Uparrow,\Downarrow}, \label{Sigma} \\
	|\Xi^{o}\rangle^{\Uparrow,\Downarrow} &=& \frac{1}{\sqrt{2}} |s ~\mathfrak{s}(us)\rangle^{\Uparrow,\Downarrow} +\frac{1}{\sqrt{6}} |s  ~\mathfrak{a}(us)\rangle^{\Uparrow,\Downarrow} -\frac{1}{\sqrt{3}} |u  ~\mathfrak{a}(ss)\rangle ^{\Uparrow,\Downarrow}. \label{Xi} 
\end{eqnarray}
In analogy with $p$, depending on their quark flavor structure, the flavor decomposition of different hyperons can be considered. The two particle Fock state expansion for the case of scalar \cite{Brodsky:2000ii} and axial-vector diquark \cite{Bacchetta:2008af} is respectively expressed as
\begin{eqnarray} 
|q \mathfrak{s}\rangle^{\Uparrow , \Downarrow}_{X_{q \mathfrak{s}}} &=& \int \frac{dx~ d^2\bfk}{16 \pi^3 \sqrt{x(1-x)}} ~ \sum_{\lambda_q} \psi^{\Uparrow, \Downarrow X_{q\mathfrak{s}}}_{\lambda_q} (x,\bfk) ~\big|\lambda_q,x P^+_X,\bfk \rangle , \\
|q \mathfrak{a} \rangle^{\Uparrow , \Downarrow}_{X_{q \mathfrak{a}}} &=& \int \frac{dx~ d^2\bfk}{16 \pi^3 \sqrt{x(1-x)}} \sum_{\lambda_q} \sum_{\lambda_a} \psi^{\Uparrow, \Downarrow X_{q \mathfrak{a}}}_{\lambda_q \lambda_{\mathfrak{a}}} (x,\bfk) ~\big|\lambda_q \lambda_{\mathfrak{a}},x P^+_X,\bfk \rangle ,
\end{eqnarray}
where $\lambda_q$ (= $\uparrow,\downarrow$) and $\lambda_{\mathfrak{a}}$ $(= \pm,0)$ denote the helicity of an active quark and axial-vector diquark. The transverse and longitudinal polarization is expressed by the symbols $\pm$ and $0$, respectively. In general, the LCWFs for a scalar diquark can be defined as
\begin{eqnarray}
	\psi^{\lambda_X X_{q \mathfrak{s}}}_{\lambda_q} (x,\bfk) = \sqrt{\frac{k^+}{(P_X-k)^+}} \frac{1}{k^2-m^2_q} \bar{u} (k,\lambda_q) ~\mathcal{Y}_{\mathfrak{s}}~ U(P_X,\lambda_X),
	\label{ScalarWfn}
\end{eqnarray}
where the scalar vertex has the form $\mathcal{Y}_{\mathfrak{s}}=ig^X_{\mathfrak{s}}(k^2) \bf{1}$ and $m_q$ corresponds to the mass of an active quark. $u(k,\lambda_q)$ and $U(P,\lambda_X)$ represent the spin-$\frac{1}{2}$ Dirac spinors for an active quark $q$ and baryon respectively. In the similar vein, the LCWFs for an axial-vector diquark can be defined as
\begin{eqnarray}
	\psi^{\lambda_X X_{q \mathfrak{a}}}_{\lambda_q \lambda_{\mathfrak{a}}}  (x,\bfk) = \sqrt{\frac{k^+}{(P_X-k)^+}} \frac{1}{k^2-m^2_q} \bar{u} (k,\lambda_q) \epsilon^\ast_\mu (P_X-k,\lambda_{\mathfrak{a}}) \cdot \mathcal{Y}_{\mathfrak{a}}^\mu~ U(P_X,\lambda_X) .
	\label{VectorWfn}
\end{eqnarray}
The four-vector polarization of spin-$1$ diquark with total momentum $(P_X-k)$ is denoted by the term $\epsilon_\mu (P_X-k,\lambda_{\mathfrak{a}})$. To uphold the spin sum rule, helicities must adhere to the constraint $\lambda_X=\lambda_q+\lambda_D+L_z$, dictated by the angular momentum conservation. Here, the symbol $D(=\mathfrak{s},\mathfrak{a})$ denotes the diquark that can be a scalar $\mathfrak{s}$ or axial-vector $\mathfrak{a}$. The relative orbital angular momentum projection between the quark and the diquark is expressed by the quantity $L_z$. for the axial-vector diquark, vertex has the form $\mathcal{Y}_{\mathfrak{a}}^\mu=ig^X_{\mathfrak{a}}(k^2) \gamma^\mu \gamma_5 /\sqrt{2}$. For the dipolar vertex, the form factor $g^X_D$ can be expressed as
\begin{eqnarray}
	g^X_{D}(k^2)=g^X_{D} \frac{(k^2-m^2_q)(1-x)^2}{(\bfk^2+L_{X_D}^2)^2},
\end{eqnarray}
where $g^X_D$ is a coupling constant for a given diquark $D$ of a baryon $X$. From Eq. (\ref{ScalarWfn}), we obtain LCWFs for scalar diquark as
\begin{eqnarray}
	\psi^{\Uparrow X_{q \mathfrak{s}}}_\uparrow (x,\bfk) &=& \frac{m_q+x M_X}{x} \, \phi^X_{{q\mathfrak{s}}}, \nonumber \\
	\psi^{\Uparrow X_{q \mathfrak{s}}}_\downarrow (x,\bfk) &=& -\frac{k_r}{x}  \phi^X_{{q\mathfrak{s}}}, \nonumber \\
	\psi^{\Downarrow X_{q \mathfrak{s}}}_\uparrow (x,\bfk) &=& -[\psi^{\Uparrow X_{q \mathfrak{s}}}_\downarrow (x,\bfk)]^\ast, \nonumber \\
	\psi^{\Downarrow X_{q \mathfrak{s}}}_\downarrow (x,\bfk) &=& \psi^{\Uparrow X_{q \mathfrak{s}}}_\uparrow (x,\bfk).
	\label{ScalarLCWFs}
\end{eqnarray}
Similarly, from Eq. (\ref{VectorWfn}), the LCWFs for axial-vector diquark come out to be
\begin{eqnarray}
	\psi^{\Uparrow {X_{q \mathfrak{a}}}}_{\uparrow +}(x,\bfk) &=& \frac{k_l}{x(1-x)} \, \phi^X_{{q\mathfrak{a}}}, \nonumber \\
	\psi^{\Uparrow {X_{q \mathfrak{a}}}}_{\uparrow -} (x,\bfk) &=& -x \frac{k_r}{x(1-x)}  \phi^X_{{q\mathfrak{a}}}, \nonumber \\
	\psi^{\Uparrow {X_{q \mathfrak{a}}}}_{\downarrow +} (x,\bfk) &=&\frac{m_q+x M_X}{x} \, \phi^X_{{q\mathfrak{a}}}, \nonumber \\
	\psi^{\Uparrow {X_{q \mathfrak{a}}}}_{\downarrow -} (x,\bfk) &=& 0 \, , \nonumber \\
	\psi^{\Downarrow {X_{q \mathfrak{a}}}}_{\uparrow +}(x,\bfk) &=& 0 \, , \nonumber \\
	\psi^{\Downarrow {X_{q \mathfrak{a}}}}_{\uparrow -} (x,\bfk) &=& -\psi^{\Uparrow X_{q \mathfrak{a}}}_{\downarrow +} (x,\bfk), \nonumber \\
	\psi^{\Downarrow {X_{q \mathfrak{a}}}}_{\downarrow +} (x,\bfk) &=& [\psi^{\Uparrow X_{q \mathfrak{a}}}_{\uparrow -} (x,\bfk)]^\ast , \nonumber \\
	\psi^{\Downarrow {X_{q \mathfrak{a}}}}_{\downarrow -} (x,\bfk) &=& [\psi^{\Uparrow X_{q \mathfrak{a}}}_{\uparrow +}(x,\bfk)]^\ast, \nonumber \\
	\psi^{\Uparrow X_{q \mathfrak{a}}}_{\uparrow 0}(x,\bfk) &=& \frac{\bfk^2-x m_{\mathfrak{a}}^2-m_q M_X (1-x)^2}{\sqrt{2}x(1-x)m_{\mathfrak{a}}} \, \phi^X_{{q\mathfrak{a}}}, \nonumber \\
	\psi^{\Uparrow X_{q \mathfrak{a}}}_{\downarrow 0} (x,\bfk) &=& \frac{(m_q+M_X)k_r}{\sqrt{2}x m_{\mathfrak{a}}}  \phi^X_{{q\mathfrak{a}}}, \nonumber \\
	\psi^{\Downarrow X_{q \mathfrak{a}}}_{\uparrow 0} (x,\bfk) &=& [\psi^{\Uparrow X_{q \mathfrak{a}}}_{\downarrow 0} (x,\bfk)]^\ast, \nonumber \\
	\psi^{\Downarrow X_{q \mathfrak{a}}}_{\downarrow 0} (x,\bfk) &=& - \psi^{\Uparrow X_{q \mathfrak{a}}}_{\uparrow 0}(x,\bfk).
		\label{VectorLCWFs}
\end{eqnarray}
The quantities $k_r$ and $K_L$ used in above equations are defined as $k_{r(l)}=k_1\pm ik_2$. The momentum space wave function $\phi^X_{{qD(=\mathfrak{s}/\mathfrak{a})}}$ written in above mentioned LCWFs can be expressed as
\begin{eqnarray}
\phi^X_{qD} =- \frac{g^X_{D}}{\sqrt{1-x}} \frac{x (1-x)^2}{[\bfk^2+L_{X_D}^2]^2} \, ,
\label{MSWF}
\end{eqnarray}
with $L^2_{X_D}=x m^2_D+(1-x) (\Lambda^X_{qD})^2-x(1-x)M^2_X$. Here, $m_D$ denotes the mass of a diquark $D$ and $\Lambda^X_{qD}$ (GeV) is an appropriate cut-off to avoid the singularity and satisfy the relation $m_D>M_X-\Lambda^X_{qD}$ \cite{Jakob:1997wg}.
\section{Numerical Parameters \label{secnumpar}}
The key input parameters involved in our calculations are the baryon masses $M_{X}$, quark masses $m_{q}$, and the spectator diquark mass $m_D$. Values of the cut-off parameter $\Lambda^X_{qD}$ (GeV) have been chosen to get the comparable values of tensor charges with available data and we have used  $\Lambda^p_{u\mathfrak{s}(uu)}=0.45$, $\Lambda^p_{u\mathfrak{a}(uu/ud)}=0.40$ in the units of GeV for proton, $\Lambda^{\Sigma^+}_{u\mathfrak{s}(us)}=\Lambda^{\Sigma^+}_{u\mathfrak{a}(us)}=0.55$, $\Lambda^{\Sigma^+}_{s\mathfrak{a}(uu)}=0.7$, $\Lambda^{\Xi^o}_{u\mathfrak{a}(ss)}=\Lambda^{\Xi^o}_{s\mathfrak{a}(us)}=0.55$ and $\Lambda^{\Xi^o}_{s\mathfrak{a}(us)}=0.65$ for different diquarks of hyperons. Following Ref. \cite{Jakob:1997wg} for proton, particle data group and Ref. \cite{Zhang:2016qqg} for other particles, the values of the masses of baryons are taken as $M_p=0.938$ GeV, $M_{\Sigma^+}=1.189$ GeV and $M_{\Xi^o}=1.314$ GeV, whereas the masses of  quarks and diquarks  are summarized in Table \ref{tab_qmass}. The coupling constant $g^X_D$ is absorbed while normalizing the unpolarized distributions $f_1(x)$ for all the considered hyperons as done in Ref. \cite{Kaur:2024yqr}. \par 
\begin{table}[h]
	\centering
	\begin{tabular}{|c|c|c|c|c|c|c|c|}
		\hline
		$\text{Quantity}  $~~&~~$ m_{u/d}  $~~&~~$  m_s $~~&~~$ m_{\mathfrak{s}(uu/ud)} $~~&~~$ m_{\mathfrak{a}(uu/ud)} $~~&~~$ m_{\mathfrak{s}(us/ds)} $~~&~~$ m_{\mathfrak{a}(us/ds)} $~~&~~$ m_{\mathfrak{a}(ss)} $ \\
		\hline
		$\text{Values (GeV)} $~~&~~$ 0.36 $~~&~~$ 0.48 $~~&~~$ 0.60 $~~&~~$ 0.80 $~~&~~$ 0.75 $~~&~~$ 0.95 $~~&~~$ 1.10 $ \\
		\hline
	\end{tabular}
	\caption{Masses of quarks and their diquark systems used in the present calculations.}
	\label{tab_qmass} 
\end{table}
\section{Chiral-odd GPDs \label{SecGPDs}}
The nonforward matrix elements of lightlike correlation functions of the tensor current are termed as chiral-odd GPDs and can be defined as
\begin{eqnarray}
	F_{\lambda_X \lambda_X^\prime}=\frac{1}{2} \int \frac{dz^-}{2\pi} e^{i \bar{x}\bar{P}_X^+z^-} \bigg\langle P_X^\prime,\lambda^\prime_X \bigg| \bar{\psi}\bigg(-\frac{z}{2}\bigg) \sigma^{+j} \gamma_5 \psi \bigg(\frac{z}{2} \bigg) \bigg|P_X,\lambda_X \bigg\rangle \bigg|_{z^+=0, \bfz=0},
\end{eqnarray}
where $j(=1,2)$ corresponds to the transverse index \cite{Diehl:2001pm, Dahiya:2007mt, Kumar:2015yta} and $\bar{P}_X$ represents the average baryon momentum. The gauge link appearing in between the initial and final state quark field operators ($\psi(z/2)$ and $\bar{\psi}(-z/2)$, respectively) becomes unity as we choose the light-cone gauge $A^+=0$. Further, the chiral-odd GPDs can be parameterized as
\begin{eqnarray}
	F_{\lambda_X \lambda_X^\prime} &=& \frac{1}{2P_X^+} \bar{U}(P_X^\prime,\lambda^\prime_X) \bigg[H_T^{X_q}(x,\zeta,-t) \sigma^{+i} \gamma_5 + \tilde{H}_T^{X_q}(x,\zeta,-t) \frac{\epsilon^{+j \alpha \beta} \Delta_\alpha \bar{P}_{X\beta}}{M^2_X} \nonumber \\
	&+& E_T^{X_q}(x,\zeta,-t) \frac{\epsilon^{+j \alpha \beta} \Delta_\alpha \gamma_\beta}{2M_X} + \tilde{E}_T^{X_q}(x,\zeta,-t) \frac{\epsilon^{+j \alpha \beta} \bar{P}_{X\alpha} \gamma_\beta}{M_X} \bigg] U(P_X,\lambda_X).
	\label{Corr1}
\end{eqnarray}
Depending on the various helicity configurations of the baryon and its constituent valence quarks, the chiral-odd GPDs can be correlated to the following matrix elements
\begin{eqnarray}
	A_{\lambda_X^\prime \uparrow,\lambda_X \downarrow} &=& \int \frac{dz^-}{2\pi} e^{i \bar{x} \bar{P}_X^+ z^-} \langle P_X^\prime,\lambda^\prime_X|\mathcal{O}_{\uparrow,\downarrow}(z) | P_X,\lambda_X \rangle |_{z^+=0, \bfz=0}, \nonumber \\
	A_{\lambda_X^\prime \downarrow,\lambda_X \uparrow} &=& \int \frac{dz^-}{2\pi} e^{i \bar{x} \bar{P}_X^+ z^-} \langle P_X^\prime,\lambda^\prime_X|\mathcal{O}_{\downarrow,\uparrow}(z) | P_X,\lambda_X \rangle |_{z^+=0, \bfz=0},
	\label{Corr2}
\end{eqnarray}
where the operators have the form
\begin{eqnarray}
	\mathcal{O}_{\uparrow,\downarrow} &=& \frac{i}{4} \bar{\psi} \sigma^{+1}(1-\gamma_5) \psi, \nonumber \\
	\mathcal{O}_{\downarrow,\uparrow} &=& -\frac{i}{4} \bar{\psi} \sigma^{+1}(1+\gamma_5) \psi.
\end{eqnarray}
On solving the quark-quark correlator, one can explicitly get the following relations
\begin{eqnarray}
	H_T^{X_q}(x,\zeta,-t) &=& \frac{T_1^{X_q}}{\sqrt{1-\zeta^2}} - \frac{2M_X \zeta}{\epsilon \sqrt{t_0-t}(1-\zeta^2)}T_3^q,  \label{HT}\\
	E_T^{X_q}(x,\zeta,-t) &=& \frac{2M_X}{\epsilon \sqrt{t_0-t}(1-\zeta^2)} (\zeta T_3^{X_q}+T_4^{X_q}) - \frac{4M_X^2}{(t_0-t)(1-\zeta^2)\sqrt{1-\zeta^2}}(T_2^{X_q}-T_1^{X_q}), \label{ET} \nonumber \\ 
	&& \\
	\tilde{H}^{X_q}(x,\zeta,-t) &=& \frac{2M_X^2}{(t_0-t)\sqrt{1-\zeta^2}}(T_2^{X_q}-T_1^{X_q}), \label{HTtilde} \\
	\tilde{E}_T^{X_q}(x,\zeta,-t) &=& \frac{2M_X}{\epsilon \sqrt{t_0-t}(1-\zeta^2)} (\zeta T_4^{X_q}+T_3^{X_q}) - \frac{4M_X^2}{(t_0-t)(1-\zeta^2)\sqrt{1-\zeta^2}}(T_2^{X_q}-T_1^{X_q}), \nonumber \\ 
	&& \label{ETtilde}
\end{eqnarray}
where $t_0$ represents the minimum value of $t$ and $\epsilon=sgn(D^1)$, with $D^1$ as the $x$-component of $D^\alpha=\bar{P}^+\Delta^\alpha-\Delta^+\bar{P}^\alpha$ \cite{Maji:2017ill}. The matrix elements in terms of helicity basis can be expressed as $T_{1(2)}^{X_q}=A_{\Uparrow \uparrow,\Downarrow \downarrow} \pm A_{\Downarrow \uparrow,\Uparrow \downarrow}$, $T_{3(4)}^{X_q}=A_{\Uparrow \uparrow,\Uparrow \downarrow} \mp A_{\Downarrow \uparrow,\Downarrow \downarrow}$. The kinematical region $0<x<1$ at zero skewness ($\zeta=0$) corresponds to the situation in which an active quark emission nd reabsorption takes place with the same longitudinal momentum fraction $x$. Henceforth, we will present GPDs without incorporating the skewness parameter as our results focus solely on chiral-odd GPDs  and the observables that can be derived from them in the absence of skewness. The matrix elements in the overlap form of LCWFs for the case of scalar diquark are given as 
\begin{eqnarray}
	T_{1(2)}^{X_{q \mathfrak{s}}} &=& \int \frac{d^2\bfk}{16 \pi^3} \big[\psi^{\Uparrow X_{q \mathfrak{s}}}_\uparrow (x,\bfk^\prime) \psi^{\Downarrow X_{q \mathfrak{s}}}_\downarrow (x,\bfk) \pm \psi^{\Downarrow X_{q \mathfrak{s}}}_\uparrow (x,\bfk^\prime) \psi^{\Uparrow X_{q \mathfrak{s}}}_\downarrow (x,\bfk) \big],  \\
	T_{3(4)}^{X_{q \mathfrak{s}}} &=& \int \frac{d^2\bfk}{16 \pi^3} \big[\psi^{\Uparrow X_{q \mathfrak{s}}}_\uparrow (x,\bfk^\prime) \psi^{\Uparrow X_{q \mathfrak{s}}}_\downarrow (x,\bfk) \mp \psi^{\Downarrow X_{q \mathfrak{s}}}_\uparrow (x,\bfk^\prime) \psi^{\Downarrow X_{q \mathfrak{s}}}_\downarrow (x,\bfk) \big].
\end{eqnarray}
For the case of axial-vector diquarks, the matrix elements in the overlap form of LCWFs are defined as
\begin{eqnarray}
	T_{1(2)}^{X_{q \mathfrak{a}}} &=& \int \frac{d^2\bfk}{16 \pi^3} \big[ \big\{ \psi^{\Uparrow X_{q \mathfrak{a}}}_{\uparrow +} (x,\bfk^\prime) \psi^{\Downarrow X_{q \mathfrak{a}}}_{\downarrow +} (x,\bfk) + \psi^{\Uparrow X_{q \mathfrak{a}}}_{\uparrow -} (x,\bfk^\prime) \psi^{\Downarrow X_{q \mathfrak{a}}}_{\downarrow -} (x,\bfk) \nonumber \\ &+& \psi^{\Uparrow X_{q \mathfrak{a}}}_{\uparrow 0} (x,\bfk^\prime) \psi^{\Downarrow X_{q \mathfrak{a}}}_{\downarrow 0} (x,\bfk) \big\}  \pm \big\{ \psi^{\Downarrow X_{q \mathfrak{a}}}_{\uparrow +} (x,\bfk^\prime) \psi^{\Uparrow X_{q \mathfrak{a}}}_{\downarrow +} (x,\bfk) \nonumber \\ &+& \psi^{\Downarrow X_{q \mathfrak{a}}}_{\uparrow -} (x,\bfk^\prime) \psi^{\Uparrow X_{q \mathfrak{a}}}_{\downarrow -} (x,\bfk) + \psi^{\Downarrow X_{q \mathfrak{a}}}_{\uparrow 0} (x,\bfk^\prime) \psi^{\Uparrow X_{q \mathfrak{a}}}_{\downarrow 0} (x,\bfk) \big\} \big], \\
	T_{3(4)}^{X_{q \mathfrak{a}}} &=& \int \frac{d^2\bfk}{16 \pi^3} \big[ \big\{ \psi^{\Uparrow X_{q \mathfrak{a}}}_{\uparrow +} (x,\bfk^\prime) \psi^{\Uparrow X_{q \mathfrak{a}}}_{\downarrow +} (x,\bfk) + \psi^{\Uparrow X_{q \mathfrak{a}}}_{\uparrow -} (x,\bfk^\prime) \psi^{\Uparrow X_{q \mathfrak{a}}}_{\downarrow -} (x,\bfk) \nonumber \\ &+& \psi^{\Uparrow X_{q \mathfrak{a}}}_{\uparrow 0} (x,\bfk^\prime) \psi^{\Uparrow X_{q \mathfrak{a}}}_{\downarrow 0} (x,\bfk) \} \mp \big\{ \psi^{\Downarrow X_{q \mathfrak{a}}}_{\uparrow +} (x,\bfk^\prime) \psi^{\Downarrow X_{q \mathfrak{a}}}_{\downarrow +} (x,\bfk) \nonumber \\
	&+& \psi^{\Downarrow X_{q \mathfrak{a}}}_{\uparrow -} (x,\bfk^\prime) \psi^{\Downarrow X_{q \mathfrak{a}}}_{\downarrow -} (x,\bfk) + \psi^{\Downarrow X_{q \mathfrak{a}}}_{\uparrow 0} (x,\bfk^\prime) \psi^{\Downarrow X_{q \mathfrak{a}}}_{\downarrow 0} (x,\bfk) \big\} \big].
\end{eqnarray} 
\begin{figure*}
	\centering
	\begin{minipage}[c]{0.98\textwidth}
		(a)\includegraphics[width=7.0cm]{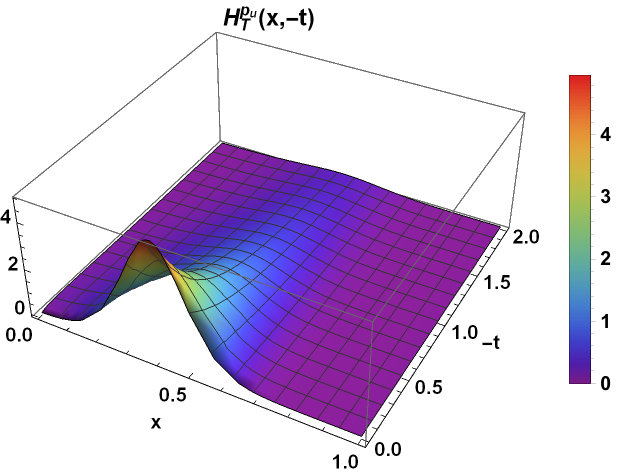}
		\hspace{0.03cm}
		(b)\includegraphics[width=7.0cm]{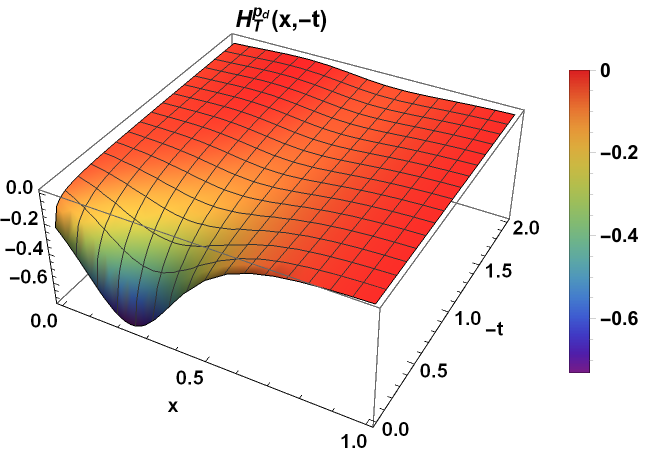}
		\hspace{0.03cm}
		(c)\includegraphics[width=7.0cm]{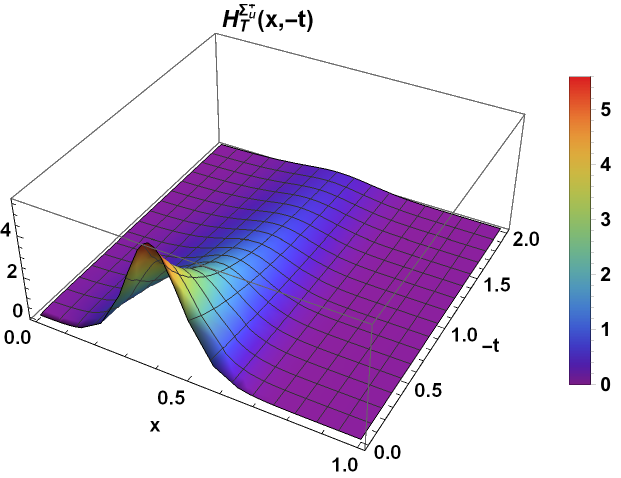}
		\hspace{0.03cm}
		(d)\includegraphics[width=7.0cm]{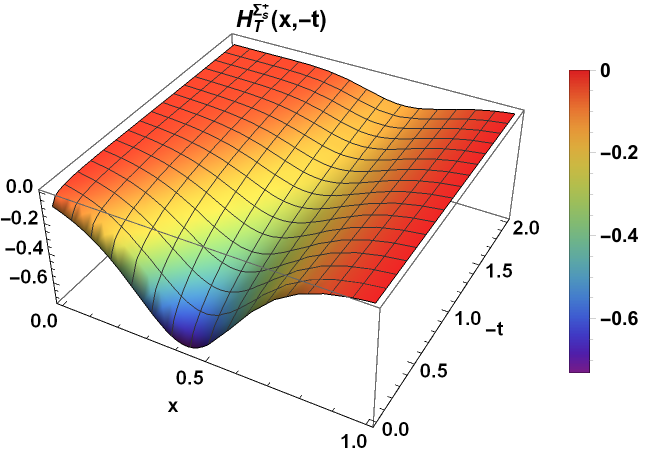}	 
		\hspace{0.03cm}	
		(e)\includegraphics[width=7.0cm]{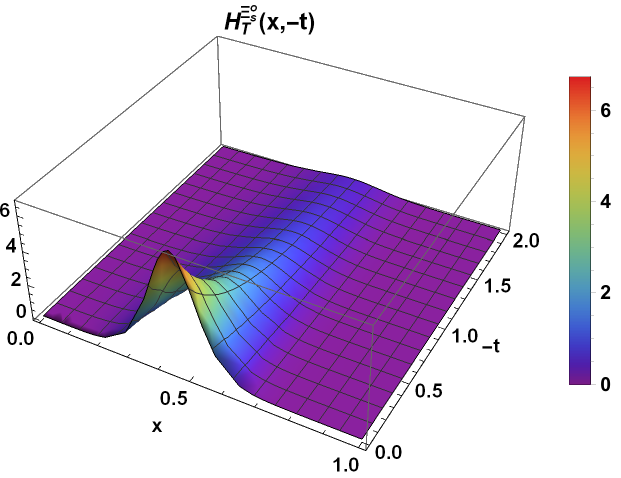}
		\hspace{0.03cm}
		(f)\includegraphics[width=7.0cm]{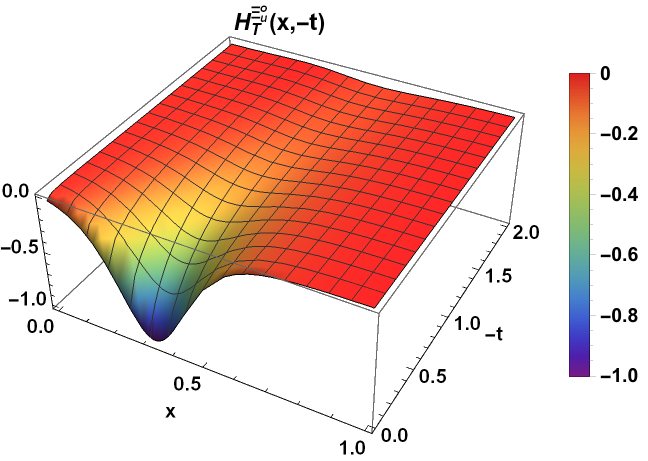}	 
		\hspace{0.03cm}	
	\end{minipage}
	\caption{\label{fig1HT} (Color online) Chiral odd GPDs $H_T^{X_q}(x,-t)$ as a function of longitudinal momentum fraction $x$ and transverse momentum transfer $-t$ (GeV$^2$) for all the constituent quark flavors of $p$, $\Sigma^+$ and $\Xi^o$.}
\end{figure*} 
In order to study the dependence of the GPDs on the longitudinal momentum fraction and transverse momentum transfer, in  Fig. \ref{fig1HT}, the 3D distributions of GPD $H_T^{X_q}$ are presented as a function of longitudinal momentum fraction $x$ and transverse momentum transfer $-t$ (GeV$^2$) for all the constituent quark flavors of $p$, $\Sigma^+$ and $\Xi^0$. According to the flavor decomposition scheme mentioned in Eq. (\ref{FlavorDecomposition}), the quark flavor distributions in the left panel incorporate contributions from both the scalar and axial-vector diquarks, while the right panel involves distributions arising solely from axial-vector diquark contribution. All the distributions show an intense peak when there is no transverse momentum transfer throughout the process, i.e., $t=0$. With the increase in the value of transverse momentum transfer, the magnitude of the distributions drops down with the movement of a peak of distributions on higher values of $x$. On comparing the distributions of the left panel, it has been found that, as a function of $-t$, the distributions of $\Sigma^+$ and $\Xi^0$ fall off comparatively slower than $p$. At $t=0$, GPD $H_T^{X_q}$ reduces to the $h_1(x)$ transversity distribution, which corresponds to the probability of finding an active quark with spin-polarized along the transverse spin of a polarized baryon minus the probability of finding it polarized oppositely. 

\begin{figure*}
	\centering
	\begin{minipage}[c]{0.98\textwidth}
		(a)\includegraphics[width=7.0cm]{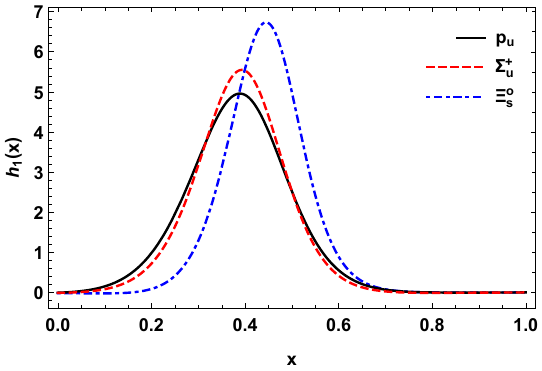}
		\hspace{0.03cm}
		(b)\includegraphics[width=7.0cm]{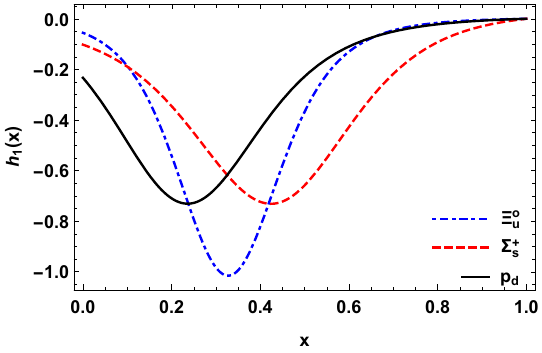}
		\hspace{0.03cm}
	\end{minipage}
	\caption{\label{fig2h1} (Color online) Comparison of transversity distributions $h_1(x)$ as a function of longitudinal momentum fraction $x$ among all the constituent quark flavors of $p$, $\Sigma^+$ and $\Xi^o$.}
\end{figure*}

A comparative analysis among $p_u$, $\Sigma^+_u$ and $\Xi^o_s$ is presented in Fig. \ref{fig2h1}(a), which involves the contributions from both scalar and axial-vector diquarks. The results indicate that the peak of transversity distribution is located at relatively higher values of the longitudinal momentum fraction $x$ for hyperons. This implies that the transverse polarization of an active quark flavor of hyperon is prominent when it carries a larger momentum fraction of the parent hyperon momentum. The distribution magnitude of both hyperons $\Sigma^+$ and $\Xi^o$ is observed to be more than that of $p$. A comparative analysis of 3D distributions arising only from axial-vector diquark contributions in the right panel of Fig. \ref{fig1HT} show negative distributions as a function of $x$ and $-t$. Among hyperons, the distribution for $\Xi^o_u$ is found to be narrower than $\Sigma^+_s$. Quantitatively, as a function of $-t$, distributions are found to fall faster for the case of $p_d$ and $56\%$ change has been found when $-t$ reaches 0.2 GeV from zero. For the same value of $-t$, the distributions of $\Sigma_s^+$ and $\Xi^o_u$ decrease by $29\%$ and $45\%$, respectively. At $t=0$, the results of transversity distribution show the peak at significantly higher values of $x$ for the case of hyperons as compared to $p$ in Fig. \ref{fig2h1}(b). The likelihood of getting a transversely polarized active quark flavor of $\Xi^o$ is higher as their magnitudes are much more than $p$ and $\Sigma^+$. As expected, $s$ quark (being heavier) is found to carry more longitudinal momentum fraction $x$ than its partner $u$ quark flavor within the same hyperon. Specifically, for the case of $p$, the results for both quark flavors are found to be consistent qualitatively with Refs. \cite{Tezgin:2024tfh, Pasquini:2005dk, Kaur:2023lun} as the magnitude of distribution is found to be more for the case of $u$ quark flavor than $d$. However, the dependence over the range of $x$ is found to be completely model dependent.

\begin{figure*}
	\centering
	\begin{minipage}[c]{0.98\textwidth}
		(a)\includegraphics[width=7.0cm]{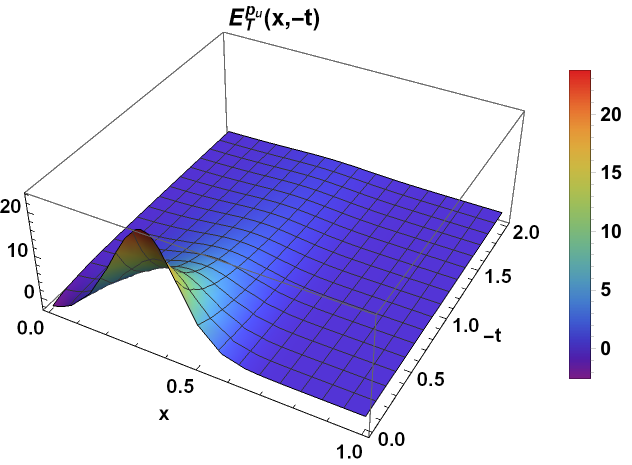}
		\hspace{0.03cm}
		(b)\includegraphics[width=7.0cm]{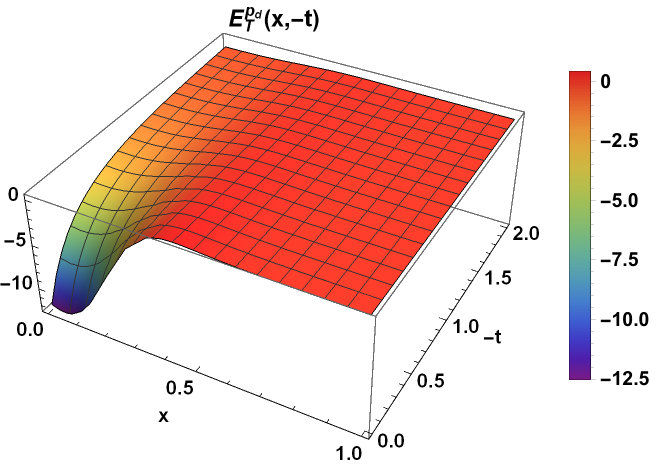}
		\hspace{0.03cm}	 
		(c)\includegraphics[width=7.0cm]{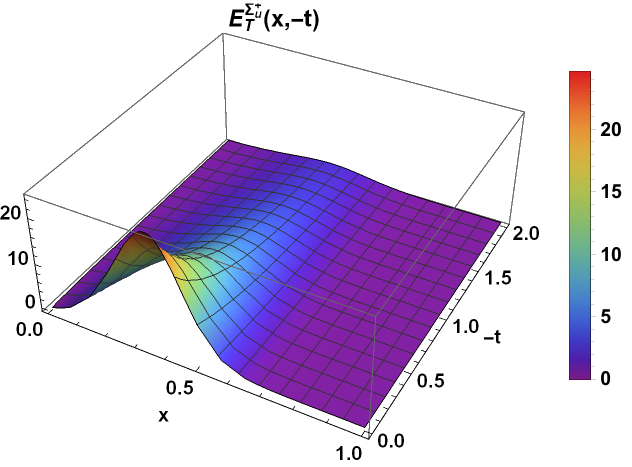}
		\hspace{0.03cm}
		(d)\includegraphics[width=7.0cm]{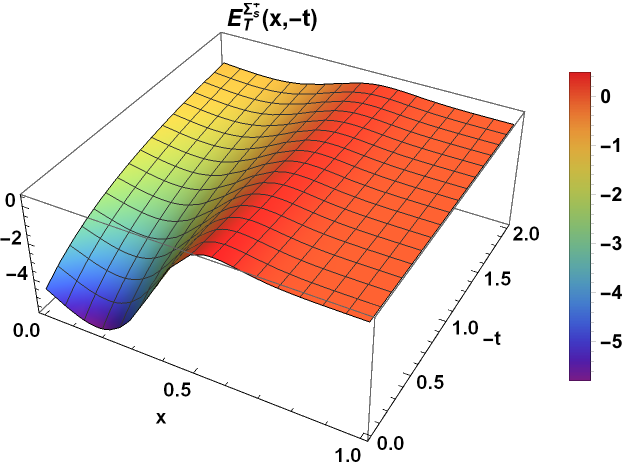}
		\hspace{0.03cm}	
		(e)\includegraphics[width=7.0cm]{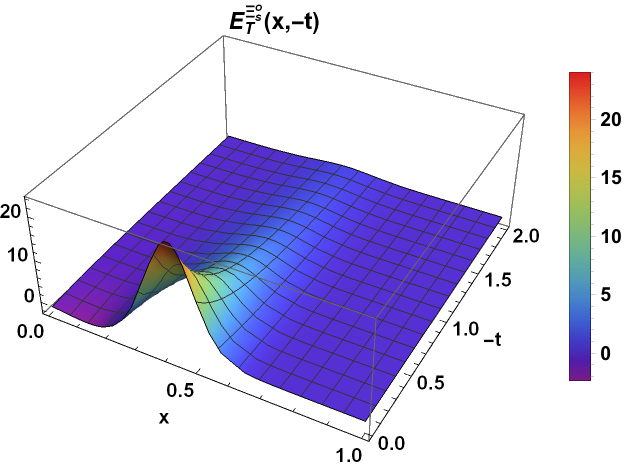}
		\hspace{0.03cm}
		(f)\includegraphics[width=7.0cm]{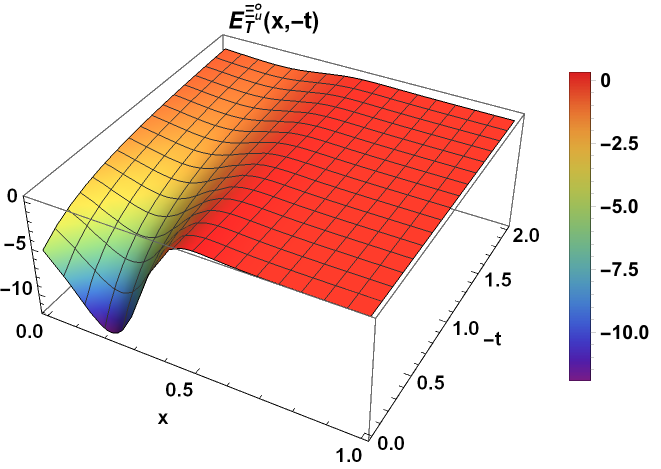}
		\hspace{0.03cm}	
	\end{minipage}
	\caption{\label{fig3ET} (Color online)  Chiral odd GPDs $E_T^{X_q}(x,-t)$ as a function of longitudinal momentum fraction $x$ and transverse momentum transfer $-t$ (GeV$^2$) for all the constituent quark flavors of $p$, $\Sigma^+$ and $\Xi^o$.}
\end{figure*}

The 3D distributions of GPD $E_T^{X_q}$ as a function of longitudinal momentum fraction $x$ and transverse momentum transfer $-t$ (GeV$^2$) for all the constituent quark flavors of $p$, $\Sigma^+$ and $\Xi^0$ are shown in Fig. \ref{fig3ET}. The left panel displays quark flavor distributions incorporating contributions from both scalar and axial-vector diquarks, whereas the right panel solely accounts for distributions arising from an axial-vector diquark contribution. For the case of $u$ constituent quark of $p$, the distribution is predominantly localized within the range $0<x<0.6$ at $t=0$. This characteristic dependence of the GPD $E_T^{p_u}$ over $x$ has also been observed in Ref. \cite{Pasquini:2005dk}. However, Refs. \cite{Kaur:2023lun, Tezgin:2024tfh} show a finite value of the distribution at $x=0$. On comparing the distributions presented in the left panel, we found that $u$ and $s$ quark flavors of $\Sigma^+$ and $\Xi^o$ have comparatively more magnitude of the distributions, along with their peaks at higher values of $x$, as compared to $u$ quark flavor of $p$ at $t=0$. Between $u$ and $s$ quark flavors of hyperons, $\Xi^o$ is observed to have a narrower distribution. The $d$ quark flavor of $p$ is found to have distribution even at a smaller region of $x$ than $u$ quark flavor of $p$ with a peak at $x=0$, which is consistent with the result obtained in Ref. \cite{Kaur:2023lun}, but zero crossing points are not observed in our model results. On comparing the distributions presented in the right panel of Fig. \ref{fig3ET} (comprising of contribution from axial-vector diquarks only), we found that the $\Sigma_u^+$ and $\Xi_s^o$ have peaks at a finite value of $x$, instead of zero. As a function of $-t$, the magnitude of the peaks fall off comparatively faster for the case of $p$ than hyperons, and the shifting of the peak towards a larger value of $x$ is consistent for all the constituent quark flavors of baryons.

\begin{figure*}
	\centering
	\begin{minipage}[c]{0.98\textwidth}
		(a)\includegraphics[width=7.0cm]{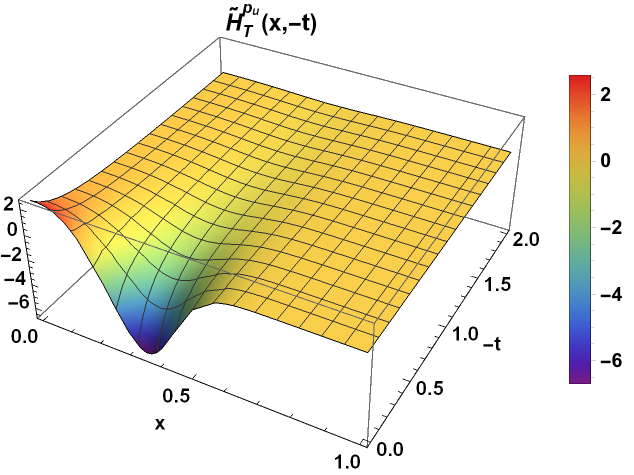}
		\hspace{0.03cm}
		(b)\includegraphics[width=7.0cm]{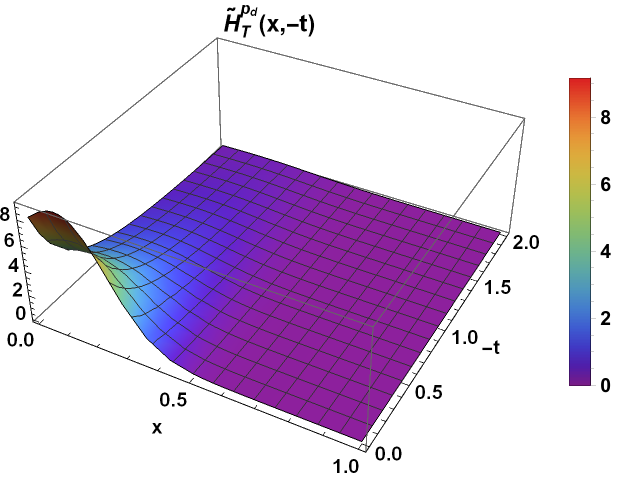}
		\hspace{0.03cm}	
		(c)\includegraphics[width=7.0cm]{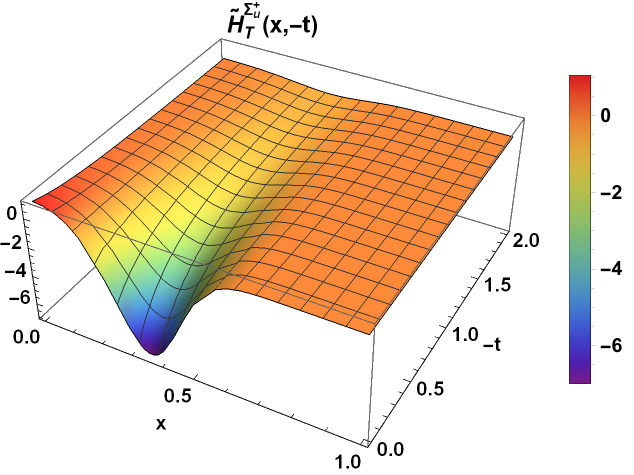}
		\hspace{0.03cm}
		(d)\includegraphics[width=7.0cm]{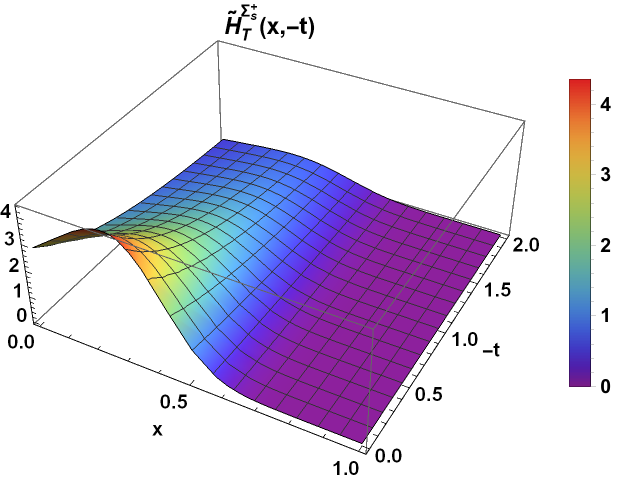} 
		\hspace{0.03cm}	
		(e)\includegraphics[width=7.0cm]{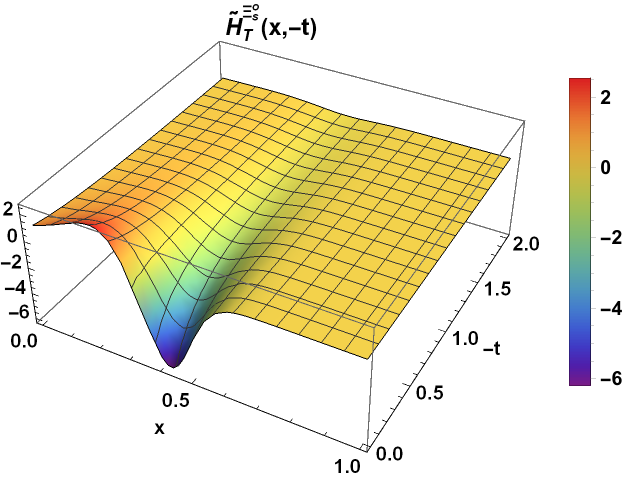}
		\hspace{0.03cm}
		(f)\includegraphics[width=7.0cm]{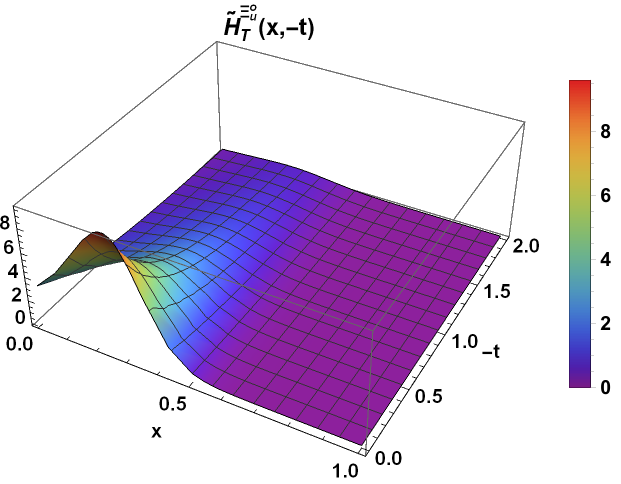} 
		\hspace{0.03cm}	
	\end{minipage}
	\caption{\label{fig4HTtlide} (Color online)  Chiral odd GPDs $\tilde{H}_T^{X_q}(x,-t)$ as a function of longitudinal momentum fraction $x$ and transverse momentum transfer $-t$ (GeV$^2$) for all the constituent quark flavors of $p$, $\Sigma^+$ and $\Xi^o$.}
\end{figure*}

In Fig. \ref{fig4HTtlide}, the 3D distributions of GPD $\tilde{H}_T^{X_q}$ as a function of longitudinal momentum fraction $x$ and transverse momentum transfer $-t$ (GeV$^2$) for all the constituent quark flavors of $p$, $\Sigma^+$ and $\Xi^0$ are illustrated. On comparing the distributions presented on the left panel, we found that for the quark flavor incorporating contributions from both scalar and axial-vector diquarks, $\tilde{H}_T^{X_q}$ presents a similar trend as shown by $H_T^{X_q}$ and $E_T^{X_q}$ as a function of $-t$. However, these distributions are negative and have a zero cross-over in the small $x$ value. Qualitatively, a similar kind of distribution concentrated on the smaller region of $x$ has also been presented in Refs. \cite{Pasquini:2005dk, Kaur:2023lun}, but without zero cross-over. The result observed are found to be completely model-dependent towards $x \rightarrow 0$. The distributions presented in the right panel solely have the contribution of axial-vector diquarks, and a comparable pattern to $E_T^{X_q}$ emerges in the distributions over $x$ with an opposite sign. The distribution $\tilde{H}_T^{p_d}$ demonstrate a resemblance to the structure presented by Ref. \cite{Kaur:2023lun}, but not with the results of Ref. \cite{Pasquini:2005dk}, particularly for $x \rightarrow 0$. On comparing it with hyperons, the distributions are found to be localized on the smaller region of $x$ for the constituent quarks of $\Xi$ than $\Sigma$ and as a function of $-t$, $\tilde{H}_T^{X_q}$ exhibits a pattern akin to $E_T^{X_q}$, but with opposite polarity.
\section{Tensor form factors \label{secFF}}
\begin{figure*}
	\centering
	\begin{minipage}[c]{0.98\textwidth}
		(a)\includegraphics[width=7.0cm]{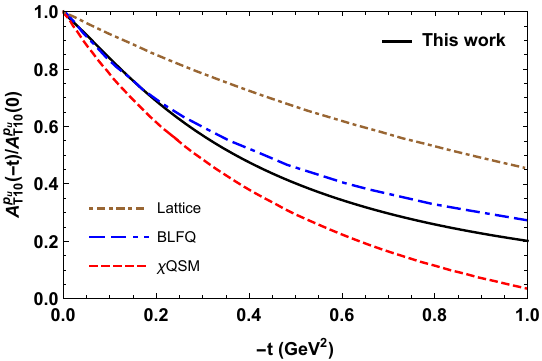}
		\hspace{0.03cm}
		(b)\includegraphics[width=7.0cm]{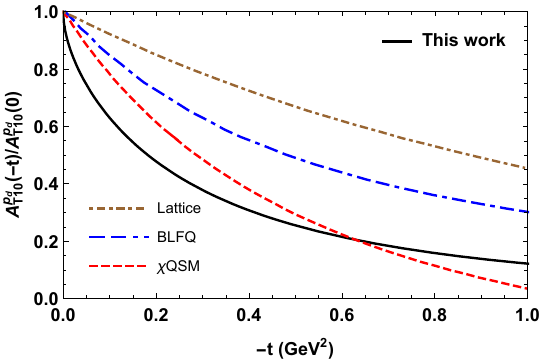}
		\hspace{0.03cm}	
	\end{minipage}
	\caption{\label{fig5AT10} (Color online) Comparison of normalized tensor form factor $A_{T10}^{p_q}(-t)$ as a function of transverse momentum transfer $-t$ (GeV$^2$) for (a) $u$ and (b) $d$ quark of a proton with lattice calculations \cite{Gockeler:2005cj}, BLFQ \cite{Kaur:2023lun} and $\chi$QSM \cite{Ledwig:2010tu} model results.}
\end{figure*}

\begin{figure*}
	\centering
	\begin{minipage}[c]{0.98\textwidth}
		(a)\includegraphics[width=7.0cm]{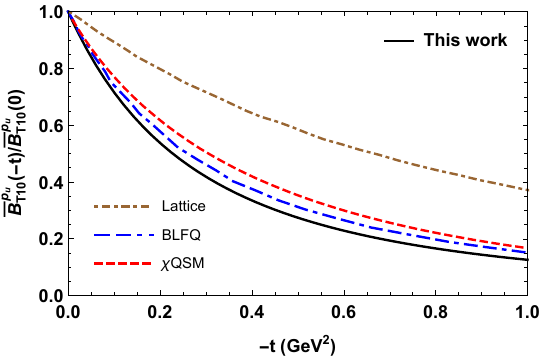}
		\hspace{0.03cm}
		(b)\includegraphics[width=7.0cm]{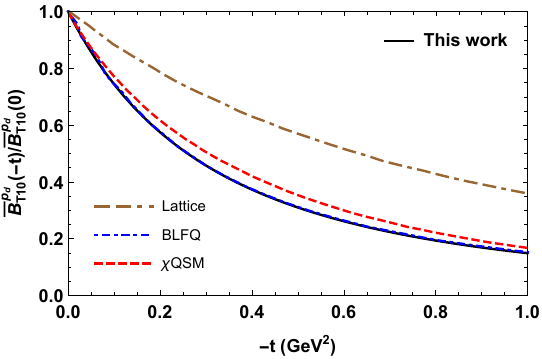}
		\hspace{0.03cm}	
	\end{minipage}
	\caption{\label{fig6BT10} (Color online) Comparison of normalized anomalous tensor magnetic form factor $B_{T10}^{p_q}(-t)$ as a function of transverse momentum transfer $-t$ (GeV$^2$) for (a) $u$ and (b) $d$ quark of a proton with lattice calculations \cite{QCDSF:2006tkx}, BLFQ \cite{Kaur:2023lun} and $\chi$QSM \cite{Ledwig:2011qw} model results.}
\end{figure*}

The moments of the GPDs, depending on their helicity configurations, provide a comprehensive insight into different form factors. Particularly, the lowest moment of chiral odd GPD $H_T^{X_q}$ encodes the tensor form factor and the combination of $\tilde{H}_T^{X_q}$ and $E_T^{X_q}$ presents the anomalous tensor magnetic form factor as
\begin{eqnarray}
	A_{T10}^{X_q}(t) &=& \int dx~H_T^{X_q}(x,-t), \label{TFF} \\
	\bar{B}_{T10}^{X_q}(t) &=& \int dx~\big(E_T^{X_q}(x,-t) + 2 \tilde{H}_T^{X_q}(x,-t) \big).
	\label{ATMM}
\end{eqnarray}
Since we are dealing with zero skewness, we have represented the GPDs in above equations as a function of remaining two variables only.
Fig. \ref{fig5AT10} presents the comparative analysis of tensor form factor for the constituent quark flavors of $p$ with lattice calculations \cite{Gockeler:2005cj}, BLFQ \cite{Kaur:2023lun} and $\chi$QSM \cite{Ledwig:2010tu} model results. The $u$ quark flavor of $p$ has a comparable trend to other model predictions as a function of $-t$ (GeV$^2$). However, a fast fall off has been observed in our model results for the case of $d$ quark flavor of $p$. For anomalous tensor magnetic form factor, the distribution is shown in Fig. \ref{fig6BT10} for both constituent quark flavors of $p$ with their comparison with lattice calculations \cite{QCDSF:2006tkx}, BLFQ \cite{Kaur:2023lun} and $\chi$QSM \cite{Ledwig:2011qw} model results. Compared to tensor form factors, anomalous tensor magnetic form factors are more consistent with the BLFQ and $\chi$QSM model results. The form factors obtained from lattice QCD are computed at a higher scale and include heavy pion mass, as a result of which the resulting form factors are straight.

\begin{figure*}
	\centering
	\begin{minipage}[c]{0.98\textwidth}
		(a)\includegraphics[width=7.0cm]{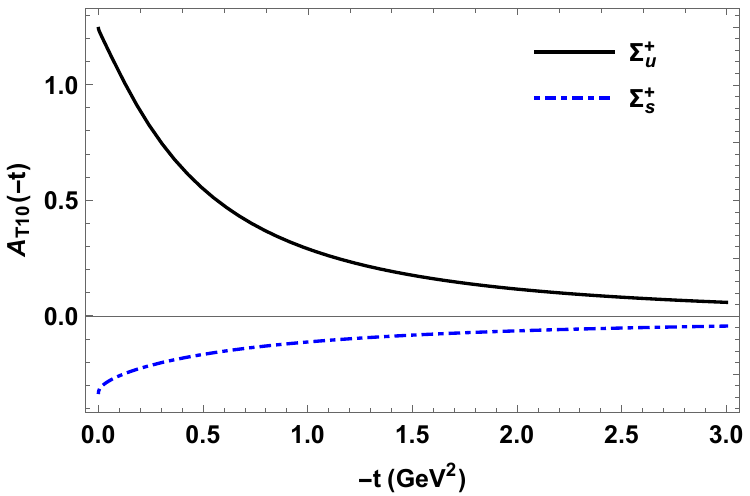}
		\hspace{0.03cm}
		(b)\includegraphics[width=7.0cm]{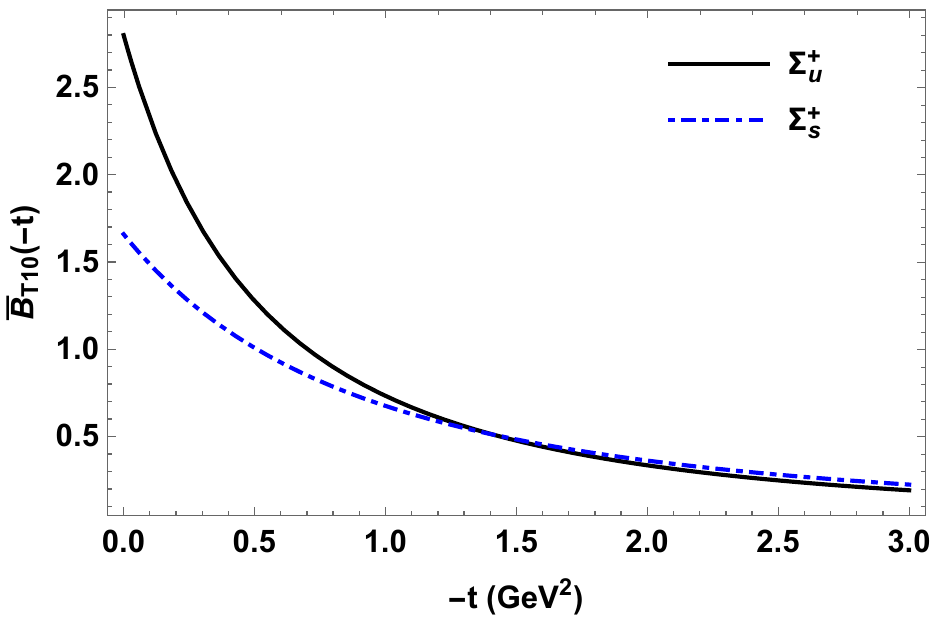}
		\hspace{0.03cm}	 \\
		(c)\includegraphics[width=7.0cm]{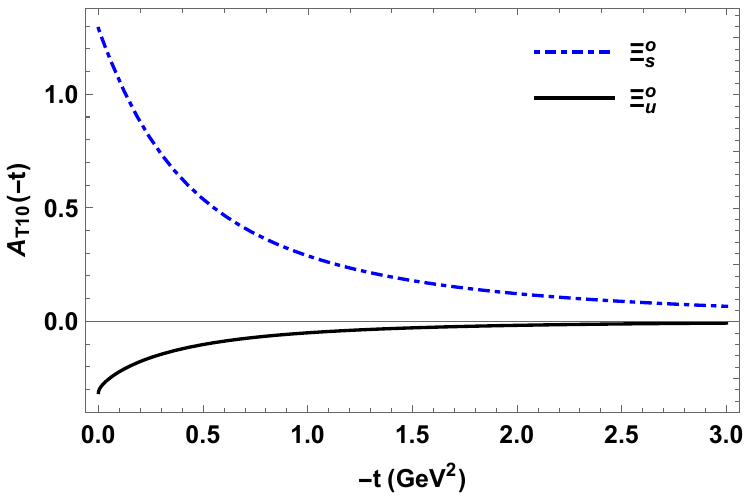}
		\hspace{0.03cm}
		(d)\includegraphics[width=7.0cm]{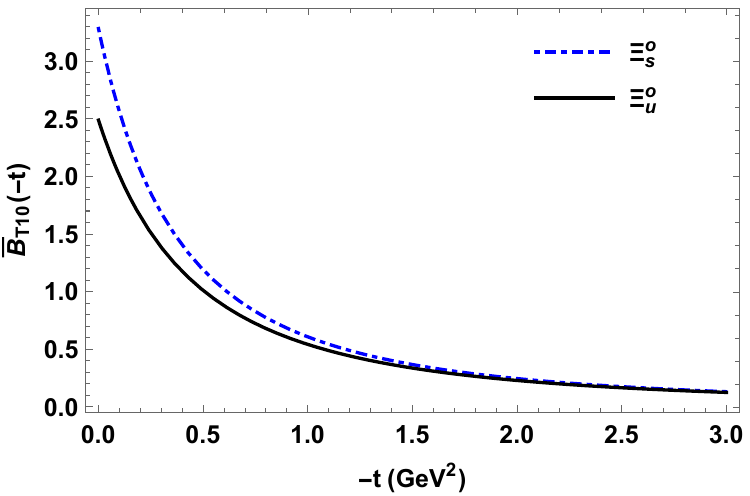}
		\hspace{0.03cm}	
	\end{minipage}
	\caption{\label{fig7} (Color online) Comparison of tensor form factor $A_{T10}^{X_q}(-t)$ (left panel) and anomalous  tensor magnetic form factor $\bar{B}_{T10}^{X_q}(-t)$ (right panel) as a function of transverse momentum transfer $-t$ (GeV$^2$) for $u$ and $s$ constituent quark flavors of hyperons.}
\end{figure*}

In Fig. \ref{fig7}, tensor form factors of $u$ and $s$ quark flavors of both hyperons are presented in subplots (a) and (c). The trend for both $\Sigma^+_u$ and $\Xi^o_s$ looks identical as both contain the contributions of scalar as well as axial-vector diquarks. A similar trend of negative distribution between $\Sigma^+_s$ and $\Xi^o_u$ has also been observed. These role reversal tensor form factors for the $u$ and $s$ quark flavors of hyperons are expected due to their analogous wave functions as expressed in Eqs. (\ref{Sigma}) and (\ref{Xi}). Further, the anomalous tensor magnetic form factor of both the constituent quark flavors of $\Sigma^+$ and $\Xi^o$ are presented in (b) and (d) subplots of Fig. \ref{fig7}, respectively. The $u$ quark contributes a larger anomalous tensor magnetic moment compared to the $s$ quark in $\Sigma^+$ hyperon and vice versa trend between quark flavors of $\Xi^o$ is observed. In both hyperons, $s$ quark flavor being a heavier one decay slowly than $u$ quark flavor. As the transverse momentum transfer increases, the difference between the quark flavors starts diminishing. 

At $t=0$, the form factors expressed in Eqs. (\ref{TFF}) and (\ref{ATMM}) can be reduced to the tensor charge $g_T^{X_q}$ and anomalous tensor magnetic moment $\kappa_T^{X_q}$, respectively. $g_T^{X_q}$ quantifies the total count of transversely polarized quark flavors $q$ in a transversely polarized baryon $X$. A comparative analysis of $g_T^{X_q}$ and $\kappa_T^{X_q}$ values from this work and other approaches is provided in Table \ref{tabComp}. Ref. \cite{Ledwig:2010tu} employs the chiral quark-soliton model to compute $g_T^{X_q}$ for octet baryons, and our results align well with their results, showing minimal deviations. For the case of proton, as an extensive work has been done, we have compared our results for $g_T^{p_{u(d)}}$ with different SU(6)-symmetric constituent quark models: the nonrelativistic (NR), the harmonic oscillator (HO) and the hypercentral (HYP) \cite{Pasquini:2005dk}, relativistic confined quark model \cite{Gutsche:2016xff}, light-front quark model (LFQM) \cite{Schmidt:1997vm}, chiral soliton model \cite{Kim:1996vk}, MIT bag model \cite{Adler:1975he}, Lattice QCD \cite{Aoki:1996pi, Abdel-Rehim:2015owa}, BLFQ \cite{Kaur:2023lun} and data analysis \cite{Anselmino:2013vqa, Kang:2015msa, Radici:2015mwa}. The values of $g_T^{p_{u(d)}}$ computed within the theoretical framework correspond well to each other than with those derived from data analysis. $\kappa_T^{X_q}$ determines the transverse spin-flavor dipole moment in an unpolarized baryon and serves a comparable role to the anomalous magnetic moment $\kappa_q^{X_q}$, computed from the Pauli form factor. To the best of our knowledge, anomalous tensor magnetic moment values of constituent quark flavor of $p$ are only available, and the comparison shows well agreement with the results of SU(6)-symmetric HO model \cite{Pasquini:2005dk} and BLFQ \cite{Kaur:2023lun}. Whereas the results of SU(6)-symmetric HYP model \cite{Pasquini:2005dk} and relativistic confined quark model \cite{Gutsche:2016xff} for both $p_u$ and $p_d$, and reggeized diquark model \cite{Goldstein:2014aja} for $p_d$ exhibit deviations.

\begin{table}[h]
	\centering
	\begin{tabular}{|c|c|c|c|c|c|c|c|}
		\hline
		\multicolumn{2}{|c|}{Baryon} & $p_u$ & $p_d$ & $ \Sigma^{+}_u $ & $ \Sigma^{+}_s $ & $ \Xi^{o}_u $ &$ \Xi^{o}_s $  \\ 
		\hline
		\multirow{15}{*}{$~~g_T^{X_q}~~$} & $~~$This work$~~$ &$~~$ 1.249 $~~$&$~~$ -0.285 $~~$&$~~$ 1.243 $~~$&$~~$ -0.331 $~~$&$~~$ -0.313 $~~$&$~~$ 1.323 \\ 
		\cline{2-8}
		& $\chi$QSM \cite{Ledwig:2010tu} & 1.08 & -0.32  & 1.08 & -0.29 & -0.32 & 1.06 \\ 
		\cline{2-8}
		& HYP model \cite{Pasquini:2005dk} & 0.97 & -0.24 & - & - & - & - \\ 
		\cline{2-8}
		& HO model \cite{Pasquini:2005dk} & 1.17 & -0.29  & - & - & - & - \\
		\cline{2-8}
		& NR model \cite{Pasquini:2005dk} & 4/3 & -1/3 & - & - & - & - \\ 
		\cline{2-8}
		& Relativistic model \cite{Gutsche:2016xff} & 1.008 & -0.247  & - & - & - & - \\
		\cline{2-8}
		& LFQM \cite{Schmidt:1997vm} & 1.167 & -0.292 & - & - & - & - \\ 
		\cline{2-8}
		& Chiral soliton model \cite{Kim:1996vk} & 1.12 & -0.42 & - & - & - & - \\ 
		\cline{2-8}
		& MIT bag model \cite{Adler:1975he} & 1.105 & -0.275  & - & - & - & - \\
		\cline{2-8}
		& Lattice QCD \cite{Aoki:1996pi} & 0.84 & -0.23 & - & - & - & - \\ 
		\cline{2-8}
		& Lattice QCD \cite{Abdel-Rehim:2015owa} & 0.791 $\pm$ 0.053 & -0.236$\pm$0.033 & - & - & - & - \\ 
		\cline{2-8}
		& BLFQ \cite{Kaur:2023lun} & 1.251 & -0.270  & - & - & - & - \\
		\cline{2-8}
		& Data analysis \cite{Anselmino:2013vqa} &$ 0.39_{-0.12}^{+18} $~~&~~$ -0.25_{-0.10}^{+30} ~~$& - & - & - & - \\ 
		\cline{2-8}
		& Data analysis \cite{Kang:2015msa} & $ 0.39_{-0.20}^{+16} $~~&~~$ -0.22_{-0.10}^{+31} ~~$ & - & - & - & - \\ 
		\cline{2-8}
		& Data analysis \cite{Radici:2015mwa} & $ 0.39\pm 0.15 $~~&~~$ -0.41\pm0.52 ~~$  & - & - & - & - \\
		\hline
		\multirow{6}{*}{$\kappa_T^{X_q}$} & This work & 3.64 & 2.41 & 3.65 & 1.658 & 2.489 & 3.284 \\ 
		\cline{2-8}
		& HYP model \cite{Pasquini:2005dk} & 1.98 & 1.17  & - & - & - & - \\ 
		\cline{2-8}
		& HO model \cite{Pasquini:2005dk} & 3.60 & 2.36  & - & - & - & - \\
		\cline{2-8}
		& BLFQ \cite{Kaur:2023lun} & 3.208 & 2.432  & - & - & - & - \\ 
		\cline{2-8}
		& Relativistic model\cite{Gutsche:2016xff} & 4.065 & 1.961  & - & - & - & - \\
		\cline{2-8}
		&  Reggeized model \cite{Goldstein:2014aja} & 3.43$\pm$0.26 & 1.37$\pm$0.34  & - & - & - & - \\
		\hline
	\end{tabular}
	\caption{Comparison of tensor charge $g_T^{X_q}$ and anomalous tensor magnetic moment $\kappa_T^{X_q}$ for all the constituent quark flavors of baryons with available data.}
	\label{tabComp}
\end{table}
\section{Summary and conclusion \label{SecCon}}
In this work, chiral-odd generalized parton distributions (GPDs) have been investigated for nucleons and the hyperons that belong to the same octet baryon of spin-$\frac{1}{2}$ hadrons. The light-cone dynamics is used to account the quark dynamics of $p$, $\Sigma^+$, and $\Xi^o$ by  treating them as two-body systems comprising a quark and a spectator diquark. A dipolar form factor for the baryon-quark-diquark vertex has been considered for computing the light-cone wave functions within the diquark spectator model. The chiral-odd GPDs are determined by solving the quark-quark correlator with a tensor current. The nonforward matrix elements of the lightlike correlation function are expressed explicitly in the overlap form of light-cone wave functions.

The results obtained for the chiral-odd GPDs $H_T$, $E_T$ and $\tilde{H}_T$ as a function of longitudinal momentum fraction $x$ and transverse momentum transfer $-t$ (GeV$^2$) at zero skewness is portrayed. 
For the case of $p$, the results are consistent with other model predictions. As a function of $-t$, the distributions corresponding to hyperons exhibit a slower decline than proton. Among constituent quark flavors of hyperons, $u$ and $s$  quark flavors of $\Sigma^+$ are found to have a slower pace of decrement than $\Xi^o$ as $-t$ increases. The characteristic property of shifting of peak towards a higher value of $x$ is also found in all the distributions with falling down as a function of $-t$. The dependence over $x$ is visualized through the transversity distributions of constituent quark flavors of considered octet baryons. The $s$ quark flavor being heavier than $u$ quark flavor in $\Sigma^+$ and $\Xi^o$ is found to carry larger values of $x$.

The tensor form factor and anomalous tensor magnetic form factors for the $u$ and $d$ quark flavors of $p$ are show consistency when compared with other models. Likewise, these form factors are computed for $u$ and $s$ quark flavors of both hyperons. As expected, the heavier $s$ quark flavor is found to decrease slower than the lighter $u$ quark flavor. Further, tensor charge and anomalous tensor magnetic moment are also computed. The results of the members of octet baryons are available for tensor charge within a theoretical framework, and our findings align well with their previous studies. Regarding the anomalous tensor magnetic moment, we have tabulated the values obtained in this work. Recent investigations by BaBar, CLEO, Belle, and BESIII of hyperon elastic and transition form factors set the stage for further studies of these hadrons. Our theoretical predictions offer insight into tensor currents in hadrons, contributing to the study of chiral-odd distributions.
\section{Acknowledgement}
H.D. would like to thank  the Science and Engineering Research Board, Anusandhan-National Research Foundation, Government of India under the scheme SERB-POWER Fellowship (Ref No. SPF/2023/000116) for financial support.

\end{document}